# AGE PREDICTION PERFORMANCE VARIES ACROSS DEEP, SUPERFICIAL, AND CEREBELLAR WHITE MATTER CONNECTIONS

*Yuxiang Wei[1,2,3], Tengfei Xue[3,4], Yogesh Rathi[3], Nikos Makris[3], Fan Zhang[3], Lauren J. O'Donnell[3]*

[1] University of Electronic Science and Technology of China, Chengdu, China
[2] University of Glasgow, UK
[3] Harvard Medical School, Boston, USA
[4] University of Sydney, Sydney, Australia

## ABSTRACT

The brain's white matter (WM) undergoes developmental and degenerative processes during the human lifespan. To investigate the relationship between WM anatomical regions and age, we study diffusion magnetic resonance imaging tractography that is finely parcellated into fiber clusters in the deep, superficial, and cerebellar WM. We propose a deep-learning-based age prediction model that leverages large convolutional kernels and inverted bottlenecks. We improve performance using novel discrete multi-faceted mix data augmentation and a novel prior-knowledge-based loss function that encourages age predictions in the expected range. We study a dataset of 965 healthy young adults (22-37 years) derived from the Human Connectome Project (HCP). Experimental results demonstrate that the proposed model achieves a mean absolute error of 2.59 years and outperforms compared methods. We find that the deep WM is the most informative for age prediction in this cohort, while the superficial WM is the least informative. Overall, the most predictive WM tracts are the thalamo-frontal tract from the deep WM and the intracerebellar input and Purkinje tract from the cerebellar WM.

## 1. INTRODUCTION

The human brain's white matter (WM) is crucial in neurological function, neurodevelopment, and brain disease [1]. Diffusion MRI (dMRI) tractography [2] can enable the study of the white matter tracts across the lifespan, including neurodevelopment and neurodegeneration [3]–[5]. Many informative quantitative features, such as fractional anisotropy (FA), axial, radial, and mean diffusivities (AD, RD, MD), can be extracted from tractography [6]. Using these features, many studies have demonstrated a link between changes in WM microstructure and aging using statistical methods [4], [7], [8]. Recently, several studies have applied machine learning or deep learning models to predict age using dMRI tractography. For instance, Mwangi et al. [9] applied relevance vector regression (RVR) to select features from the WM (including FA, MD, RD, and AD) and predict age. They included 207 subjects with an age range of 4-85 years and achieved a mean absolute error (MAE) between 5 and 10 years within different age groups and diffusion measures. Richard et al. [10] studied 24 diffusion features, including FA, MD, AD, and RD, of 612 subjects with an age range of 18-87 years. They employed extreme gradient boosting (xgboost) to make predictions and achieved 7.28 years MAE. Chen et al. [11] proposed a transfer learning pipeline to train a fully connected neural network. The method was evaluated on three datasets and achieved a MAE of 4.78 years on 616 subjects with age range of 18-88 years, 5.35 years on 405 subjects with age range of 18-92 years, and 5.64 years on 176 subjects with age range of 20-82 years. The diffusion features studied included AD, RD, MD, FA, volume ratio, generalized FA, non-Gaussianity (NG), orthogonal NG, and parallel NG.

However, these studies have not yet investigated which anatomical connections of the WM may be most predictive of age in different age ranges. Furthermore, these studies demonstrate that age prediction is a challenging problem and developing methods with increased accuracy (by reducing MAE, for example) is important. For these purposes, this work has two main contributions. First, we propose a novel deep learning framework for improved age prediction based on dMRI data. Second, we apply this framework to investigate the age predictivity of different anatomical white matter regions in a large dataset of healthy young adults. Overall, we show that the proposed method outperforms compared state-of-the-art methods and enables identification of connections that are most predictive of age in this young adult population.

## 2. METHODOLOGY

### 2.1. Dataset

The dataset included in this paper is derived from the "1200 Subjects Data Release" dataset from the Human Connectome Project (HCP) Young Adult Study [12], [13]. The data processing pipeline for tractography and tract parcellation is described in [14]. Briefly, to compute whole-brain tractography, the two-tensor unscented Kalman filter (UKF) [15] method is employed via the ukftractography package of

SlicerDMRI [16], [17]. Following a recursive estimation order, the UKF method fits a mixture model of two tensors to the diffusion data while tracking fibers. The first tensor models the traced tract, and the second tensor models fibers crossing the tract. UKF tractography is highly consistent across ages, health conditions, and image acquisitions [18], [19]. Afterward, tractography parcellation is performed based on a neuroanatomist-curated WM atlas [19]–[21]. Fiber clusters are categorized into three groups: left- and right-hemisphere and commissural/decussating. It should be noted that the clusters in the left and right hemispheres can be matched one-to-one since the atlas clustering is conducted bilaterally. Moreover, further categorization can be made for each cluster.

For this study, the tractography parcellation contains 42 tracts, with 953 clusters as we discard the false positive and unclassified clusters according to [19], The tracts are categorized into three groups: deep, superficial, and cerebellar, according to their anatomical regions. Note that the deep WM contains 29 tracts with 467 clusters, the superficial WM contains eight tracts with 388 clusters, and the cerebellar WM contains five tracts with 98 clusters.

For each cluster, we compute its FA and MD for Tensor1 and Tensor2, and record their max, min, median, mean, and variance. Therefore, given the tensor $t \in \{t_1, t_2\}$, we extract the statistical measures $s \in \{max, min, median, mean, var\}$ of diffusion properties $p \in \{FA, MD\}$, resulting in 20 features per cluster. Moreover, we also include the total number of points and streamlines. Hence, each cluster possesses 22 features. These features provide comprehensive information for our deep learning model. Note that the data is normalized by calculating the L2 norm of each cluster's features so that all features have the same numeric range (0 to 1). Also note that unlike image data that has pixel neighborhood context, our input data lacks contextual information about the relationship between different features (i.e., no neighborhood or time-series relationships are encoded in our data).

### 2.2. Data augmentation

It is a generally accepted notion that training on a large dataset can result in better deep learning models [22]. For models trained on a limited dataset, the overfitting problem commonly occurs, which severely impairs the performance. To improve the generalization ability, numerous data augmentation methods have been published for image data and time-series data [23]. Cutmix [24] is a straightforward algorithm that is widely applied with excellent performance on image data. In this study, we propose a discrete multi-faceted mix (DMF-Mix) algorithm based on Cutmix to augment data that lacks contextual information.

As presented in Algorithm 1, we first calculate the mixing ratio λ from a beta distribution according to [25]. For each subject in the input $x$, we randomly select the clusters to be mixed. Then for the features $x_{i,j} \in \mathbb{R}^{1 \times L}$ of each cluster, we randomly select the features $I_{feature}$ that are to be replaced and $\hat{I}_{feature}$ that are used to replace. $I_{feature}$ are replaced by $\hat{I}_{feature}$, which produces the mixed data $\hat{x}$. It should be noted that the proposed method does not enlarge the dataset's size. For each of the original samples, features of clusters are randomly mixed to generate the augmented data.

Like Cutout [26], DMF-Mix mixes the features and forces the model to take more of the whole features into consideration rather than focusing on a small set of features that may not always imply age for some subjects, thereby improving its generalizing ability. Meanwhile, unlike Cutmix [24] that selects complete regions of input and mixes them, we randomly select features from some clusters within each subject and replace them with other features. This induces more random alterations to the input and effectively prevents over-fitting. Furthermore, we perform the mix within each subject, which is demonstrated to be better than mixing across subjects.

---

Algorithm 1: discrete multi-faceted mix (DMF-Mix)

Input: $x \in \mathbb{R}^{N \times H \times L}$ where $N$ is the number of subjects, $H$ is the number of clusters, and $L$ is the number of features per cluster.
Output: augmented data $\hat{x} \in \mathbb{R}^{N \times H \times L}$.

1 Generate mixing ratio $\lambda = Beta(1,1)$;
2 $I_{cluster}$ = randomly select $H \times \sqrt{1-\lambda}$ clusters;
3 **For** $x_i$ in $x$ **do**   // loop through each subject
4    **For** $j$ in $I_{cluster}$ **do**   // loop through each cluster
5      $I_{feature} \leftarrow$ randomly select $L \times \sqrt{1-\lambda}$ features from $x_{i,j}$;
6      $\hat{I}_{feature} \leftarrow$ randomly select $L \times \sqrt{1-\lambda}$ features from $x_{i,j}$;
7      $\hat{x}_{i,j} \leftarrow \hat{x}_{i,j}[\hat{I}_{feature}] = x_{i,j}[I_{feature}]$;
8 **Return** $\hat{x}$

---

### 2.3. The age prediction model (AP-model)

To extract crucial features from the augmented data, we design a CNN that includes inverted bottleneck and residual connections. After extracting useful features via CNN, the output feature map is flattened and projected to the latent space using a feed-forward network, which generates the prediction. The architecture of the model is shown in **Fig. 1**.

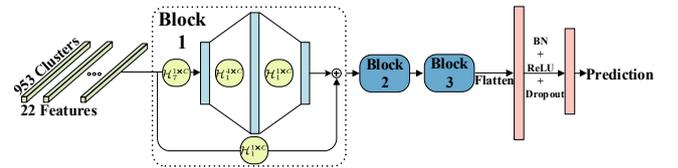

Fig. 1 The proposed AP-model. We stack three identical convolution blocks to perceive key information. Note that $\mathcal{H}_k^{\alpha C}$ donates a 1D convolution layer with $\alpha \times C$ channels and kernel length of $k$, followed by batch-normalization, ReLU, and Dropout.

As introduced in Section 2.1, the input data has 965 subjects in total, each with 953 clusters and the corresponding 22 features. We regard the clusters as the input channels and the features as the length. For the extraction of information, we designed a convolution block with three critical features following the key concepts of state-of-the-art works [27]–[29]: large kernel size, inverted bottleneck, and residual connection. This joint mechanism enables our framework to perceive information within multiple scales while ensuring computational efficiency.

As shown in **Fig. 1**, for each block, we first apply a depth-wise (DW) convolution with $C$ channels and a kernel length of 7. As the input channel (number of clusters) is 953, we set $C$ to be 1024 as 1024 is the least multiple of 2 that is bigger than 953. Instead of using length of 3 kernels (a popular choice), we choose larger convolutional kernels (length of 7) as they offer a larger receptive field, which allows convolution layers to gather more information from a large region [30]. As a result, $\mathcal{H}_7^C$ can integrate information across multiple features for each cluster. It should be noted that $\mathcal{H}_7^C$ also includes a batch-normalization layer, a ReLU, and a Dropout after each Conv. These add non-linearity to the learning and mitigate over-fitting.

Followed by the first conv layer, we use two point-wise (PW) conv layers: $\mathcal{H}_1^{4C}$ to scale the channel with a ratio of 4 and then use $\mathcal{H}_1^C$ to recover. This inverted bottleneck does a non-linear expansion over the original feature space, thus enriching the information for training. In addition, instead of implementing the standard conv, we implement depth-wise separable conv by combining the DW length 7 conv ($\mathcal{H}_7^C$) and PW length 1 conv ($\mathcal{H}_1^{4C}$). This can facilitate the extraction of features: instead of mapping cross-channel (cross-cluster) and cross-feature correlations simultaneously like the standard conv, the depth-wise separable conv firstly looks at the cross-feature correlations by the DW conv, then focuses on the correlations between clusters by the PW conv. As a result, it decouples the mapping of cross-cluster and cross-feature correlations and is shown to be better at extracting useful features [32].

Finally, a residual connection with $\mathcal{H}_1^C$ is employed to facilitate back-propagation during training and allows our model to be deeper. Three blocks are stacked to generate a feature map that contains fine-grained information and is subsequently flattened to a 1D vector. A linear feed-forward network is employed to map the vector and make prediction.

### 2.4. The Priori-Loss function

For age prediction, the subjects' ages of the dataset are within a limited scale, ranging from 22 to 37. Therefore, we incorporate this as a priori knowledge into the loss function to give a harsher penalty to any prediction result that is out of this range, as in Eq. 1.

$$\mathscr{L} = \begin{cases} (P-T)^2 & 22 \leqslant P \leqslant 37 \\ (P-T)^2 \times S & otherwise \end{cases} \quad (1)$$

Here $P$ denotes the prediction result, $T$ denotes the ground truth, and $S$ denotes a scale factor. Commonly we set $S$ to be 2 so that any unexpected output (out of the range of 22 to 37) can be further avoided.

### 3. EXPERIMENTAL RESULTS

#### 3.1. Experimental settings

We evaluate the proposed model based on Pytorch 1.12.1, with Nvidia A100. We perform a 5-fold cross validation to ensure fairness during evaluation. The optimizer we choose is AdamW, with a weight decay of 0.05 and a learning rate of 0.01. To facilitate the convergence, we employ the cosine annealing learning rate scheduler. Additionally, we apply the proposed Priori-Loss as the loss function.

TABLE I. ABLATION STUDY FOR DMF-MIX

| Methods | Specification | MAE | RMSE |
|---|---|---|---|
| baseline | | 2.63 | 3.24 |
| mixup | | 2.64 | 3.24 |
| cutmix | AS | 2.59 | 3.21 |
| | WS | 2.68 | 3.23 |
| DMF-Mix | AS + F | 2.65 | 3.26 |
| | AS + C | 2.74 | 3.30 |
| | AS + FC | 2.61 | 3.21 |
| | WS + F | 2.62 | 3.27 |
| | WS + C | 2.59 | 3.21 |
| | **WS + FC** | **2.59** | **3.20** |

#### 3.2. Results on age prediction

*3.2.1. Ablation study*

We conduct an ablation study to demonstrate the effectiveness of the proposed DMF-Mix and Priori-Loss. For DMF-Mix, it randomly mixes some features of clusters within each subject. However, there are several design choices to make: performing the mixing across subjects, mixing some features of all clusters, or mixing all features of some clusters. Furthermore, we also test the performance of the classic Cutmix when applying within each subject. The results are presented in Table 1. The last row (with specification WS + FC) is our DMF-Mix. Here we include two evaluation metrics: mean absolute error (MAE) and root mean squared error (RMSE). Note that baseline denotes if no data augmentation method is used, while AS denotes mixing across subjects, WS denotes within subject, F denotes mix selected features across all clusters, C denotes mix all features within selected clusters, and FC denotes select a subset of the clusters and randomly pick features for mixing. Also note that the comparison is performed based on the whole WM.

From the table, WS has better performance than AS. For instance, the MAE for WS+C is 0.15 years lower than AS+C. This may be because the diffusion measures of subjects vary with age, so mixing them with subjects of different ages could confuse the model.

To demonstrate the effectiveness of the proposed Priori-Loss, we train the final model from the last row of Table I (DMF-Mix with WS+FC) and we replace the Priori-Loss with the MSE loss. Performance decreases with the MSE loss (2.66 years MAE and 3.34 RMSE).

TABLE II. COMPARE MODELS TRAINED ON DIFFERENT WM REGIONS

|  | MAE | | | |
| --- | --- | --- | --- | --- |
|  | ALL | DWM | SWM | CWM |
| CNN | 2.73 | 2.74 | 2.93 | 2.92 |
| RVR | 2.74 | 2.78 | 3.03 | 2.98 |
| **AP-model** | **2.59** | **2.69** | **2.95** | **2.89** |

### 3.2.2. Comparison with state-of-the-art methods

We compare the proposed model with other methods that predict age based on dMRI: which include: a 1-D CNN [33] and a relevance vector regressor (RVR) with a rational-quadratic kernel [9]. To demonstrate that different WM regions have differing age predictivities, we compare model performance on the whole white matter (ALL), deep WM (DWM), superficial WM (SWM), and cerebellar WM (CWM). As shown in Table II, the proposed AP-model outperforms other methods in all WM regions, indicating the success of the proposed DMF-Mix and Priori-Loss.

Comparing different WM regions, all models' performances on DWM are substantially better than on other WM regions. The models trained on DWM achieve similar results to the models trained on the whole WM, even though the DWM has fewer clusters (467 clusters versus 953 clusters). This suggests that DWM contains most of the valuable features for age prediction for the studied age range of 22 to 37. Interestingly, though the CWM has fewer clusters (98) than the SWM (388), all models trained on the CWM were more predictive of age than those trained on the SWM. These results may relate to regionally varying developmental trajectories of WM tracts. Individual DWM tracts reach peak FA at different ages during young adulthood [3], while FA in many SWM regions is relatively stable in young adults [34].

TABLE III. TOP PERFORMING TRACTS IN WM REGIONS

| Region | Tract Name | MAE | RMSE |
| --- | --- | --- | --- |
| DWM | thalamo-frontal (TF) - 42 | 2.86 | 3.43 |
|  | corpus callosum 2 (CC2) - 16 | 2.96 | 3.54 |
|  | striato-frontal (SF) - 30 | 2.96 | 3.55 |
|  | thalamo-parietal (TP) – 20 | 2.97 | 3.57 |
|  | corpus callosum 1 (CC1) - 3 | 2.98 | 3.57 |
| SWM | superficial-frontal (Sup-F) – 162 | 2.97 | 3.59 |
| CWM | intracerebellar input and Purkinje (Intra-CBLM-I&P) – 28 | 2.87 | 3.49 |

### 3.3. Tract-based comparison

Although the observations in the previous section provide a preliminary answer about which region is most predictive of age based on the young adult HCP dataset, we still do not know if the tracts within these regions have different predictive abilities. Therefore, we train the AP-model on each tract and compare model performance. Note that since the number of clusters in a tract is much less than that in the whole WM, we use a smaller AP-model by setting $C$ as 128. The results are shown in Table III. Here we only show tracts with relatively high performance: the top five tracts in DWM, the top tract in SWM, and the top tract in CWM. Also note that the number beside each tract name denotes the number of clusters in this tract.

Overall, the top age prediction performance is observed on thalamofrontal (TF) and intracerebellar input and Purkinje (Intra-CBLM-I-P) tracts, with MAEs of 2.86 and 2.87 years respectively. Apart from this, the number of clusters in the tract does not necessarily affect its predictivity. For instance, Sup-F has 162 clusters, yet its MAE is 2.97 years, higher than some tracts that have fewer clusters.

Our findings are somewhat consistent with previous work, where commissural and projection tracts were found to mature earlier than association connections [5]. (The top predictive DWM tracts in Table III belong to the projection and commissural categories, suggesting that tracts near the peak of their maturity are predictive of age in this study.) However, the cerebellar tracts are relatively less studied with respect to age [5]. We used an atlas with a unique cerebellar parcellation to enable the study of the medullary white matter of the cerebellum (referred to as the intracerebellar input and Purkinje tract in the atlas).

### 4. CONCLUSION

In this study, we proposed a novel deep learning method for improved age prediction, and we investigated age predictivity across anatomical connections of the white matter. Overall, we found that deep white matter connections are the most predictive of age in young adults. By applying deep learning to simultaneously study multiple white matter connections with multiple diffusion properties, this style of analysis may shed light on the nonlinear microstructure changes of the human brain during development and aging.

### 5. COMPLIANCE WITH ETHICAL STANDARDS

This research study was conducted retrospectively using human subject data made available in open access by HCP [12]. Ethical approval was not required.

### 6. ACKNOWLEDGEMENTS

We acknowledge the following NIH grants: P41EB015902, R01MH125860 and R01MH119222. F.Z. also acknowledges a BWH Radiology Research Pilot Grant Award.